# INTRANET SECURITY USING A LAN PACKET SNIFFER TO MONITOR TRAFFIC

Ogbu N. Henry[1] and Moses Adah Agana[2]

[1]Department of Computer Science, Ebonyi State University, Abakaliki, Nigeria
greatrio2016@gmail.com
[2] Department of Computer Science, University of Calabar, Nigeria
ganamos999@yahoo.com, Mosesagana@unical.edu.ng

*ABSTRACT*

*This paper was designed to provide Intranet traffic monitoring by sniffing the packets at the local Area Network (LAN) server end to provide security and control. It was implemented using five computer systems configured with static Internet Protocol (IP) addresses used in monitoring the IP traffic on the network by capturing and analyzing live packets from various sources and destinations in the network. The LAN was deployed on windows 8 with a D-link 16-port switch, category 6 Ethernet cable and other LAN devices. The IP traffics were captured and analyzed using Wireshark Version 2.0.3. Four network instructions were used in the analysis of the IP traffic and the results displayed the IP and Media Access Control (MAC) address sources and destinations of the frames, Ethernet, IP addresses, User Datagram Protocol (UDP) and Hypertext Transfer Protocol (HTTP). The outcome can aid network administrators to control Intranet access and provide security.*

## 1. INTRODUCTION

In every network, security is needed by the users. Hence, a reliable and secure connection from every computer in a network must be ensured as users communicate with each other. One of the methods to realize this is by using a packet sniffer to capture and analyze packets that run through the network.

Packet sniffing is tool used for monitoring and analyzing the network to troubleshoot and log activities. It assists in capturing all the packets on networks irrespective of the final destination of the packet. In recent times, due to network issues such as network abuse and malicious connections, the management, maintenance and monitoring of the network is important to keep the network smooth and efficient. For this purpose, a packet sniffer, sometimes called a network analyzer (formerly known as ethereal) is used. The essence is to curb malicious attacks.

With a packet sniffer, one can watch all the non-encrypted data that travel from one's computer onto the internet; this includes data that is not secured by encryption [1]. By placing a packet sniffer on a network in promiscuous mode, all data traffic travelling across the network can be viewed by the network administrator and once the raw packet data is captured, the packet sniffer analyzes it and presents it in human readable form so that the person using the software can make sense out of it.

A sniffer can be either hardware or software, which mainly intercepts and collects the local traffic. After recording the traffic, the sniffer provides the function to decode and simply analyze the content of the packets in human readable format. A packet sniffer can be used in many ways whether it is for good or bad. A packet sniffer can be used to monitor network activities. For example, when a network monitoring tool is located at one of the servers of one's Internet Service Provider (ISP), it would be potentially possible to monitor all of the network activities, such as which website is visited, what is looked at on the site, who is sent an email, what is being downloaded from a site and also what streaming events are used. The positive usage of a packet sniffer is to maintain a network so as to work normally by capturing packets, recording and analysing traffic, decrypting packets and displaying in clear text, converting data to a readable format, and showing relevant information like IP, protocol, host or server name. Though having a sniffer installed can benefit a lot in terms of network troubleshooting and network usage, the limitation of a sniffer is that it cannot read the encrypted packets [2].

An Intranet is a virtual private network (VPN) that links offices to the headquarters' internal network over a shared infrastructure using dedicated connections. Intranet VPNs allow access only an organization's employee or agent using authentication and encryption techniques. Intranet web servers therefore differ from public web servers. The public web servers do not have access to an organization's Intranet without passwords and permissions, except via hacking. Cyber criminals can go to any length to breach network securities.

A packet sniffer is a tool which captures all the packets on the network irrespective of the final destination of the packet. If it is installed in any of the nodes of the network (either as source or destination), it can be used to analyze the performance of the network or to find bottlenecks in it. Packet sniffers are of two types: active sniffers which can send data in the network and can be detected by other systems through different techniques and the passive sniffers which only collect data, but cannot be detected (e.g. wireshark). Also, the structure of a packet sniffer consists of two parts: packet analyzer which works on the application layer protocol and the packet capture (pcap) which captures packets from all other layers [2].

Packet sniffers are software tools used in monitoring network activities, akin to law enforcement investigations into criminal behaviours, monitoring the communications to and from people of interest to gather evidence about crimes that the suspects will commit [3]. Packet sniffers are installed on computers in a network, and once activated, they make copies of all network traffic packets that are sent and received by the host computer. They are used for a variety of reasons, including: as a problem solving tool to fix network problems, as a performance tool to identify bottlenecks in the network and areas where efficiency can be improved, and as a technique in security management. A network administrator can grab those packets that are passing through the network to trace any access to the network. A packet sniffer is used for detecting messages being sent and received from a network interface, detecting an error implementation in network software and collecting statistics and the network traffic.

If the network interface card (NIC) of a system is in promiscuous mode, it can take over all packets and frames it receives on a network [4]. The security threat presented by sniffers is their ability to capture all incoming and outgoing traffic, and it may seem difficult to detect these sniffing tools because they are passive in nature. The detection of such sniffing tools is however only difficult when the capturing and analyses of data is done in a SHARED environment not on a SWITCHED environment. With the information captured by the packet sniffer, administrators can identify erroneous packets, using them to pinpoint bottlenecks and help to maintain efficient network data. Some organizations see packet sniffers as internal threats. Some however see packet sniffers as just being a hacker's tool, though it can also be used for network traffic monitoring, traffic analysis and troubleshooting to avoid misuse of network by both internal and external users [5].

Slowdown in the network performance can cause some serious concern to network analysts, leading to loss in resources. Such cases are not easy to deal with due to lack of time. However, sometimes the cause of problems in a network could be due to attacks by unknown third parties that try to put the web server out of service through means of Denial of Service (DOS) attack, sending some poisoned traffic in an attempt to discover hosts to infect or simply by infecting ports with malwares. In all these cases, knowing the sources of the attacks is the first step towards taking appropriate action in achieving correct protection. A good way of achieving correct protection is by using a trusted packet sniffer such as Wireshark, formerly known as Ethereal, whose prime objective is networking troubleshooting, analysis, and networking research [6].

Packet sniffing is a method of tapping each packet as it flows across the network. It is a technique in which a user sniffs data belonging to other users of the same network. Packet sniffers can also be used for malicious purposes, or can be operated as an administrative tool. It is dependent on the user's intentions. Packet sniffers are used by the network administrators as utilities for efficient network administration. The following conclusion was drawn from the working behaviour of already existing sniffer software (i.e. wireshark):

i) Packet sniffers only give the log of data, this has to be analyzed by the network administrator to find the error or attack on the network adapter.
ii) Current systems are only able to show of packets.
iii) The limitation of protocol based traffic analysis include the fact that it is extremely time consuming to capture every packet, examine them, disassemble every one and manually take an action based on the interpretation from the analysis [7].

Cyber crime investigators and law enforcement agencies that need to monitor email during investigations likely use a sniffer designed to capture very specific kinds of traffic, knowing that sniffers simply grab network data [7].

As computer networks are increasing in their sizes very rapidly, so is the number of its users, crime rates as well as traffic flows in the networks increasing, so it is very important to monitor network traffic as well as its user's activities to keep the network smooth and efficient [4]. For a complex network, it is a very tough task to maintain and monitor the traffic because of the large amounts of data available. For this purpose, packet sniffing becomes a compendium for network management, monitoring, and ethical hacking.

When packets are transferred from source to destination, they pass through many intermediate devices. A node whose NIC is set in the promiscuous mode receives all information travelling in the network, When a switch which is already passing filtered data is used, a method to capture all data of the network is utilized [4]. Figure 1 illustrates the physical structure of a packet sniffer.

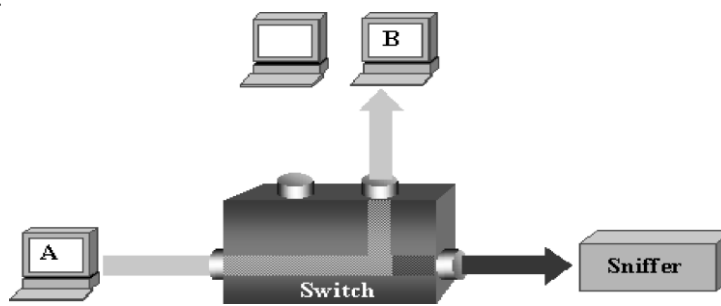

Figure 1: The physical structure of a packet sniffer [4].

## 2. PACKET SNIFFER ARCHITECTURE

Any sniffer can be divided into:

- Hardware
- Drive program
- Buffer
- Packet Analysis

Packet sniffers can be operated in both switched and non-switched environment. All business hosts are connected to the hub in the switched environment though businesses are updating their network infrastructure, replacing aging hubs with new switches. The replacement of hubs with new switches that makes switched environment widely used is to increase security. It cannot be said that packet sniffing is not possible in switched environment [8]. Network traffic analysis is the positive aspect of packet sniffing, which is analyzed by a network analyzer which can;
- Provide detailed information of activities that are going on in the network
- Test anti-malware programs and pin-point potential vulnerabilities
- Detect unusual packet characteristics
- Identify packet sources and destinations

Figure 2 shows how packets fro various sources and destinations are sniffed at the server end.

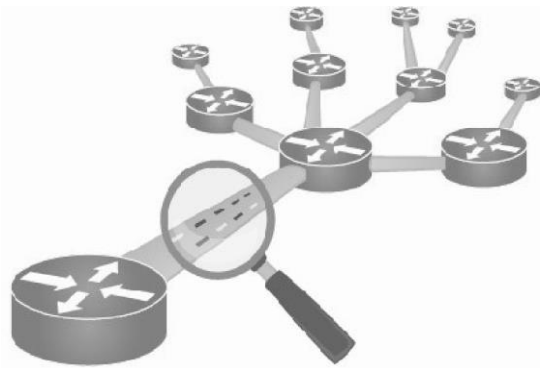

Figure 2: The Internal process of packet sniffing at server end [8].

A Packet sniffer in its simplest form captures all of the packets of data that pass through a given network interface. Typically, the sniffer will capture packets that were intended for the machine in question. However, if placed into the promiscuous mode, the packet sniffer is also capable of capturing all packets traversing the network regardless of the destination [9].

There are three sniffing methods, some methods work in switched environments while others work in non- switched environments , they include:
  i. IP- based sniffing: This is the original way of packet sniffing. It works by putting the network card into promiscuous mode and sniffing all packets matching the IP address filter. It works only in non-switched networks.
  ii. MAC-based sniffing: This method works by putting the network card into promiscuous mode and sniffing all packets matching the MAC address filter.
  iii. ARP-based sniffing: This method works a little different. It doesn't put the network card into promiscuous mode. This happens because the ARP protocol is stateless, because of this, sniffing can be done on a switched network [10].

Network sniffing describes the process of monitoring, capturing and interpreting all incoming and outgoing traffic as it flows across a network. A Packet sniffer can show you all sorts of things going on behind the scenes, including unknown communication between network devices, actual detailed error codes provided by layer-specific protocols, and even poorly designed programs going crazy [10].

## 2.1. How a Packet Sniffer Works

The packet sniffing process can be broken down into three steps: collection, conversion and analysis [10].

*Collection*: The Packet sniffer switches the selected network interface into promotion mode. In this mode, the network card can listen to all network traffics on its particular network segment to capture the raw binary data from the wire.

*Conversion*: The captured binary data is converted into a readable form. This is where most of the advanced command-line-driven packet sniffers stop. At this point, the network data is in a form that can be interpreted only on a very basic level, leaving the majority of the analysis to the end user.

*Analysis*: The packet sniffer takes the captured network data, verifies its protocol based on the information extracted, and begins its analysis of the protocol's specific features. Wireshark is one of the most popular open-source packet analyzer. Originally named Ethereal, in May 2006 the project was named Wireshark due to trademark issue.

## 3. METHODOLOGY

The system was designed using five computer systems configured with static Internet Protocol (IP) addresses used in monitoring the IP traffic on the network by capturing and analyzing live packets from various sources and destinations in the network. The LAN was deployed on windows 8 with a D-link 16-port switch, category 6 Ethernet cable and other LAN devices. The Ebonyi State University Abakaliki Intranet was used in testing the system. The IP traffics were captured and analyzed using Wireshark Version 2.0.3. Four network instructions were used in the analysis of the IP traffic. It was intended to display the IP and MAC address sources and destinations of the frames, Ethernet, IP addresses, User Datagram Protocol (UDP) and Hypertext Transfer Protocol (HTTP).

### 3.1. System Design

Figure 3 shows the physical design of the system. Cables and other devices were linked to establish communication between five offices using an access point.

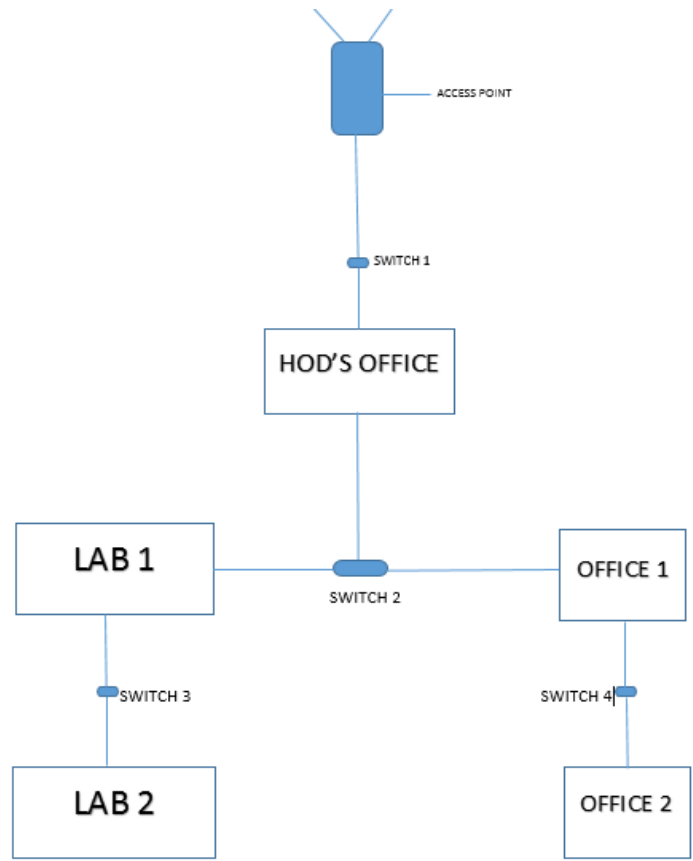

Figure 3: Physical Design of System LAN

Figure 4 shows the physical design of the Wireshark to sniff packets from five network nodes.

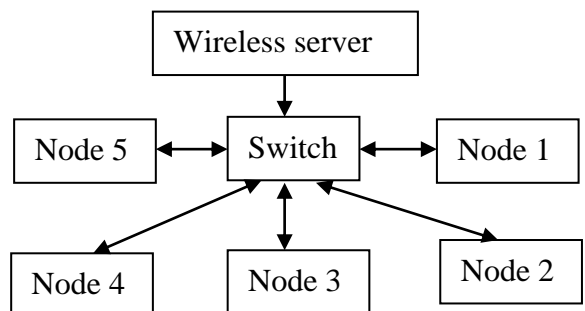

Figure 4: Physical Design of Wireshark

Before the implementation of wireshark, there must be an existing LAN. This is to ensure that different devices/systems are connected and are communicating with or without the internet. To ensure that connected devices are communicating with each other and to enable the network administrator to view and analyze activities in the network, the following steps were carried out:

- Crimping of cable
- Pinging
- File sharing

After launching the packet sniffer at the administrator's work station, it begins to capture packets on different applications such as web browsers and the File Transfer Protocol (FTP) client.

The sniffing process is as follows:

- Start
- Select capture interface
- Capture packets
- Set filter
- Stop capturing if no more streaming frames
- Save captured frames
- Analyze captured packets
- Generate report
- End.

Figure 5 shows the data flow model of the packet sniffing process. The user of the sniffed packets is the network administrator.

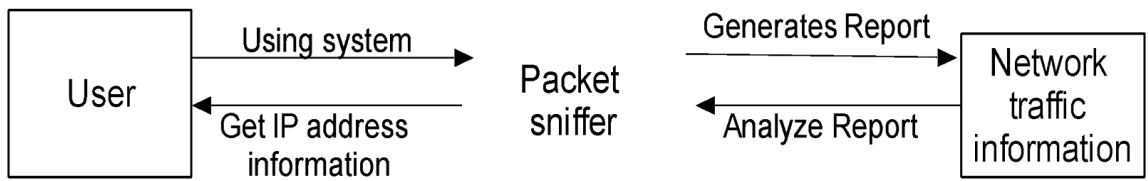

Figure 5: Data flow model of the packet sniffing process

Figure 6 shows the data flow model of the packet sniffer in the LAN.

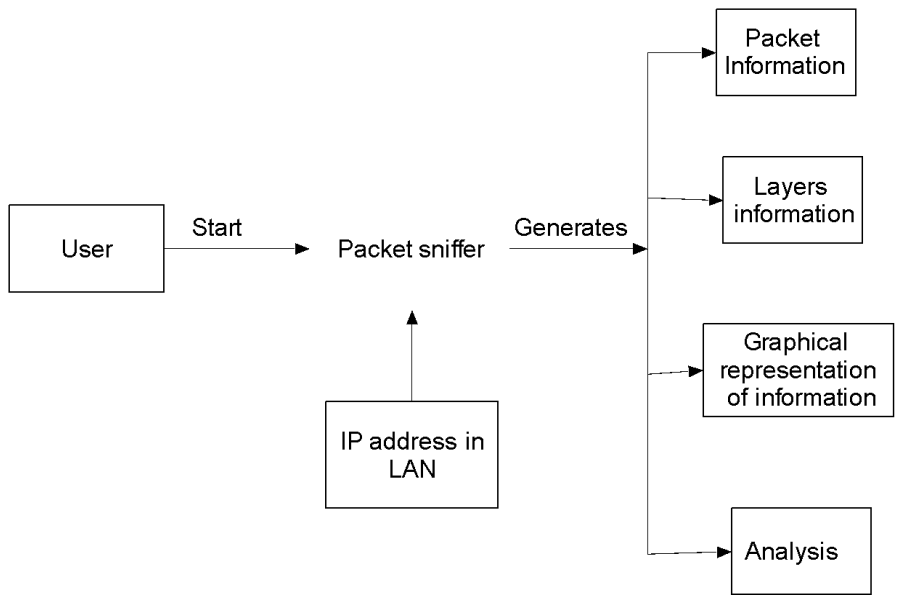

Figure 6: Data flow model of the packet sniffer in the LAN

The administrator must issue a pinging command to ascertain if there is communication between two or more systems after being connected through a bridge (switch). This is illustrated in Figure 7.

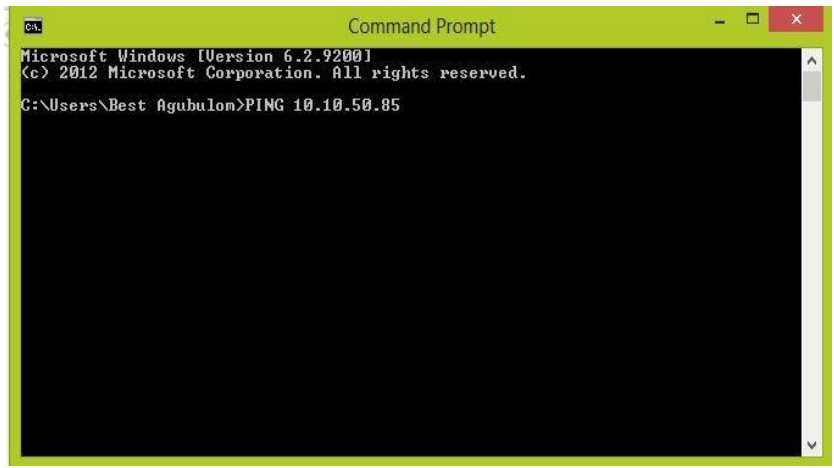

Figure 7: Pinging commamd issued by an administrator

The outcome of the pinging is shown in Figure 8.

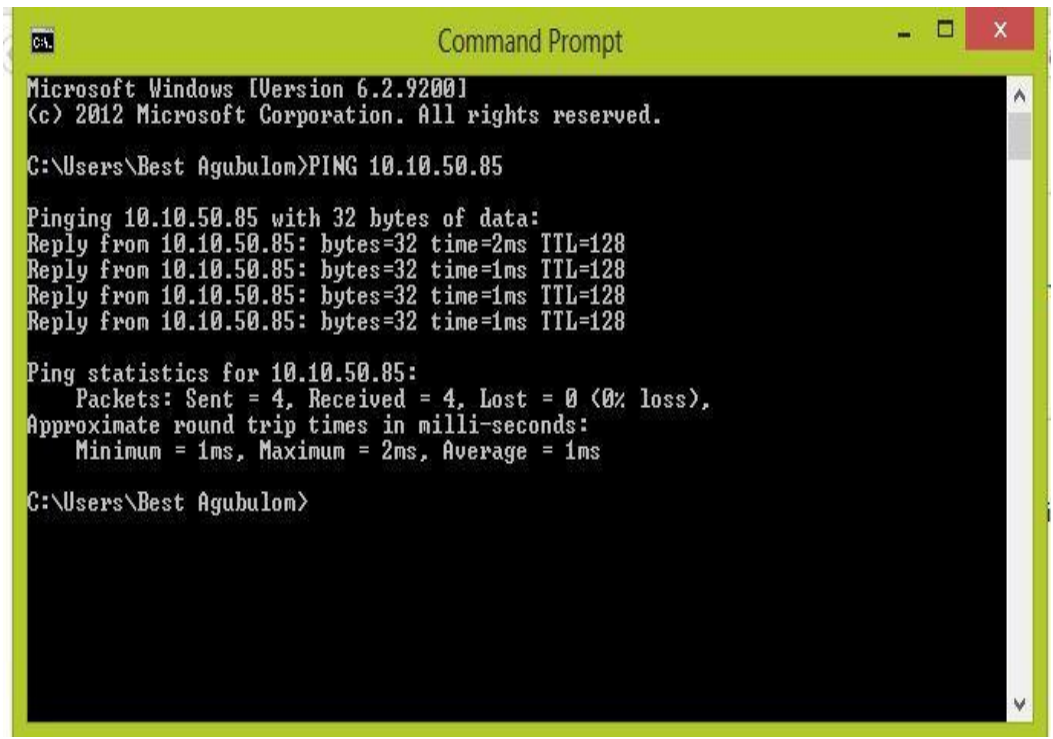

Figure 8: Result showing pinged devices and feedbacks

This window in Figure 8 confirms the connectivity of the command "ping 10.10.50.85" that was issued, which simply means there is communication between the two devices and a reply from the recipient IP address **10.10.50.85**.

## 3.2. System Requirements

The hardware required for the system include: personal computers, switches, Ethernet cables, interface cards, wireless access points, crimpers, trunks, a LAN tester, and an RJ-45 connector.

The software requirements include Wireshark (packet sniffer), Wincap (allows the network interface card to operate in 'promiscuous mode') and Windows 8 operating system.

## 3.3. Results and Discussion

The results obtained under testing the system are presented and discussed in this section. Figure 9 shows the user interface where the administrator initiates the packet sniffing process. The most common medium frequently used by network administrators to analyze network is ETHERNET.

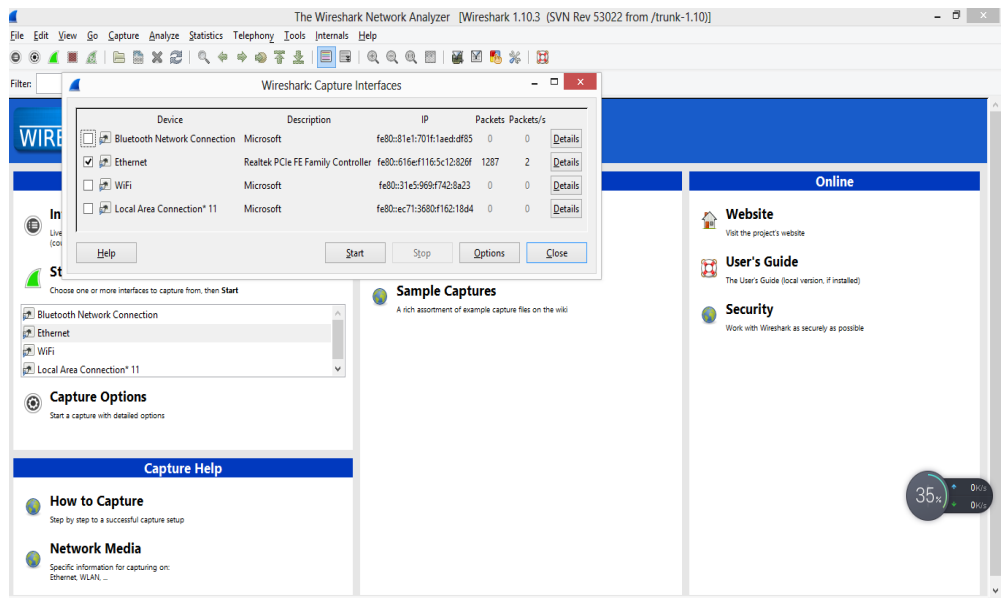

Figure 9: Wireshark user interface

The live data captured by the network sniffer showing random movements of different IP addresses of users on the network, from their sources to destination, the time, protocol, length and information on the network is illustrated in Figure 10.

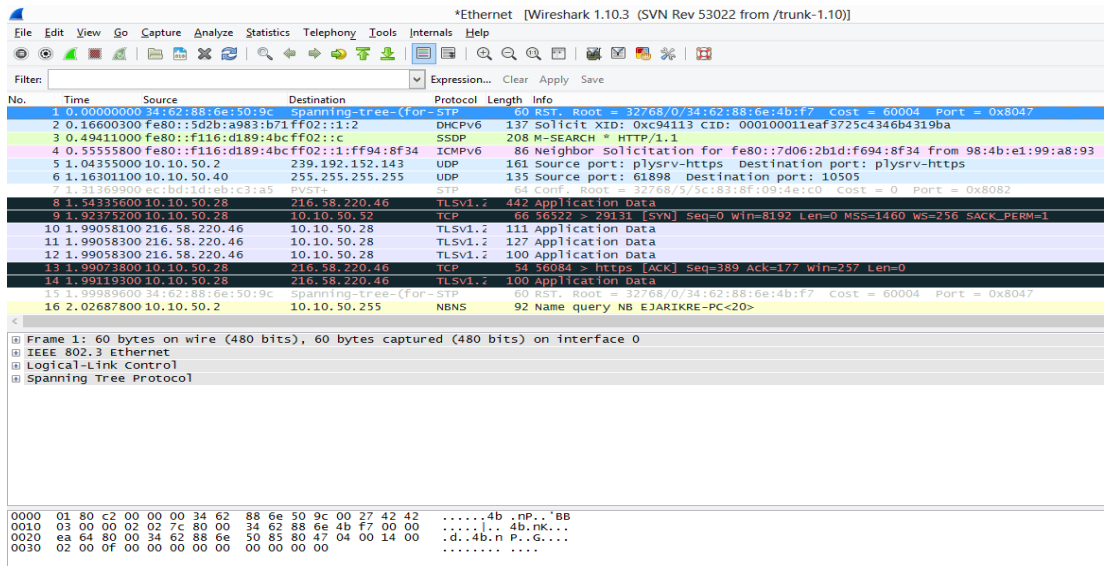

Figure 10: Captured packets

The interface in Figure 10 above shows packets running from different source IPs to destination IPs, their protocols, the time each went in, the length of data and the type of activities carried out by each user.

The captured traffic showing the frames involved, time in nanoseconds, source IP addresses and respective destination IP addresses, protocols involved, and the information depending on the activity of different hosts and length of data when the system was tested is illustrated in Figure 11.

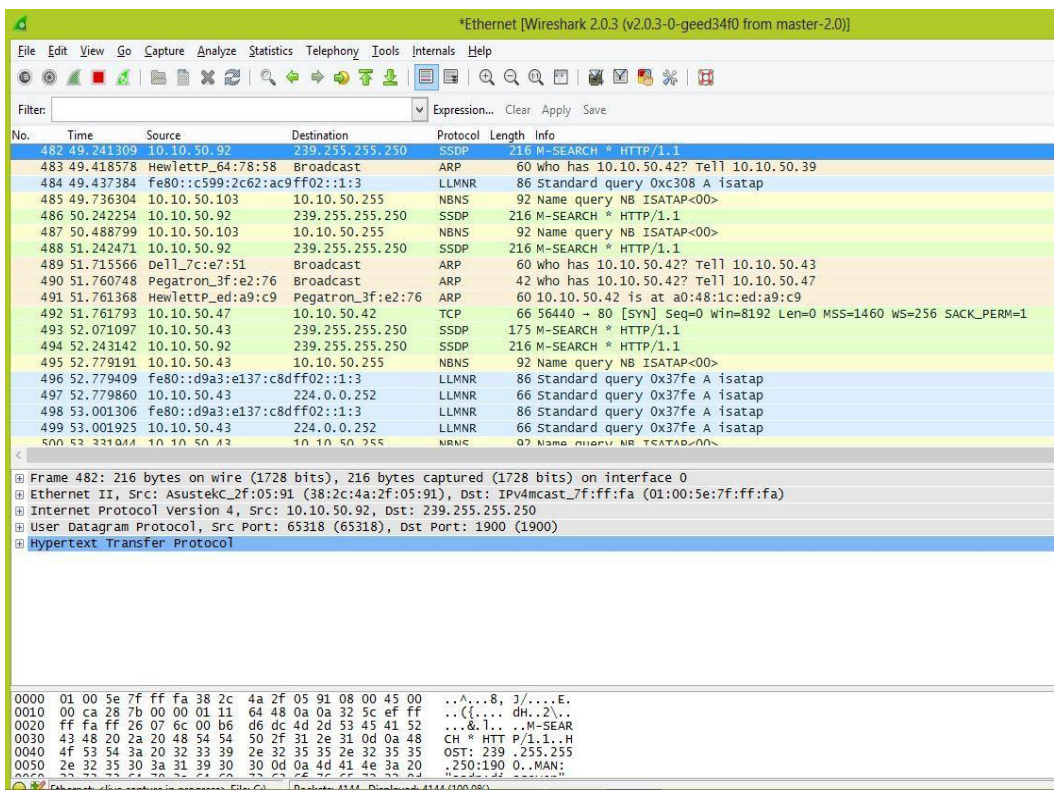

Figure 11: Captured frames

## 4. CONCLUSIONS

Network security is a sine qua non for any organizational success, considering the value of data and information transmitted via networks. The Intranet though secured and restricted to authorized members of an enterprise that owns it is still prone to cyber attacks most times.

It is necessary to watch for new user accounts or high activity on a previously low usage account, new files with novel or strange file names, accounting discrepancies, changes in file lengths or dates, attempts to write to system, data modification or deletion, denial of service, unexplained, poor system performance, anomalies, suspicious probes, suspicious browsing, inability of a user to log in due to modifications of his/her account, etc.

This necessitates the need for some sniffing software to monitor the packets and activities in an Intranet.

The results obtained from the study showed the IP and MAC address sources and destinations of the frames, Ethernet, UDP and HTTP successfully captured. These can assist network administrators to make informed decisions on possible threats that the network can be exposed to.

## ACKNOWLEDGEMENTS

The authors are grateful to the authorities of the Ebonyi State University Information and Communication Technology (ICT) Directorate for allowing us access to use the Intranet to test this work.

**Authors**

**Dr. Ogbu N. Henry** is a Lecturer at the Ebonyi State University Abakaliki. He has a Ph.D. in Computer Science (Digital Emergency Response)

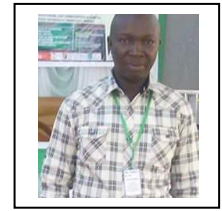

**Dr. Moses Adah Agana** is a Senior Lecturer and Head of Department of Commpuyer Science in the University of Calabar, Nigeria. He has a Ph.D. in Cyber Security.

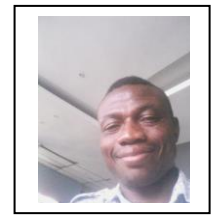